\documentclass[useAMS, usenatbib]{mn2e}

\usepackage{epsfig}

\usepackage{amssymb}

\usepackage{multirow}

\usepackage{enumitem}

\usepackage{ulem}

\usepackage{makecell}

\usepackage{rotating}
\usepackage{natbib}
\usepackage{latexsym}
\bibpunct{(}{)}{;}{a}{}{,}
\usepackage{url}

\begin{document}

\newcommand{\ledd}{%
$L_{Edd}$}

\newcommand{\IGR}{IGR~J18245--2452}

\newcommand{\Msun}{M$_\odot$}

\def\rem#1{{\bf #1}}
\def\hide#1{}

\def \aj {AJ}
\def \mnras {MNRAS}
\def \apj {ApJ}
\def \apjs {ApJS}
\def \apjl {ApJL}
\def \aap {A\&A}
\def \aapr {A\&ARv}
\def \nat {Nature}
\def \araa {ARAA}
\def \pasp {PASP}
\def \aaps {AAPS}
\def \prd {PhRvD}
\def \apss {ApSS}

\newcommand{\specialcell}[2][c]{%
  \begin{tabular}[#1]{@{}c@{}}#2\end{tabular}}

\author[Linares et al.]{
\parbox[t]{\textwidth}{
\raggedright 
Manuel Linares$^{1,2}$\thanks{manuel.linares@ntnu.no},
Barbara De Marco$^{2}$,
Rudy Wijnands$^{3}$,
Michiel van der Klis$^{3}$}
\vspace*{6pt}\\
$^1$ Institutt for Fysikk, Norwegian University of Science and Technology, Trondheim, Norway.\\
$^2$ Departament de F{\'i}sica, EEBE, Universitat Polit{\`e}cnica de Catalunya, Av. Eduard Maristany 16, E-08019 Barcelona, Spain.\\
%
%
$^3$ Anton Pannekoek Institute for Astronomy, University of
Amsterdam, Science Park 904, 1098XH Amsterdam, The Netherlands.\\
 }

\title[X-ray variability of tMSPs]{X-ray variability of transitional
  millisecond pulsars: a faint, stable and fluctuating disk}


\maketitle

\begin{abstract}

Transitional millisecond pulsars (tMSPs) have emerged in the last
decade as a unique class of neutron stars at the crossroads between
accretion- and rotation-powered phenomena.
In their (sub-luminous) accretion disk state, with X-ray luminosities
of order $10^{33}-10^{34}$~erg~s$^{-1}$, they switch rapidly between
two distinct X-ray modes: the disk-high (DH) and disk-low (DL) states.
We present a systematic {\it XMM-Newton} and {\it Chandra} analysis of the
aperiodic X-ray variability of all three currently known tMSPs, with a
main focus on their disk state and separating DH and DL modes.
We report the discovery of flat-topped broadband noise in the DH
state of two of them, with break frequencies of 2.8~mHz
(PSR~J1023+0038) and 0.86~mHz (M28-I).
We argue that the lowest frequency variability is similar to that seen
in disk-accreting X-ray binaries in the hard state, at typical
luminosities at least 2 orders of magnitude higher than tMSPs.
We find strong variability in the DH state around 1~Hz, not typical of
hard state X-ray binaries, with fractional rms amplitudes close to
30\%.
We discuss our results and use them to constrain the properties of the
accretion disk, assuming that the X-ray variability is produced by
fluctuations in mass accretion rate, and that the break frequency
corresponds to the viscous timescale at the inner edge of the disk.
In this context, we find that the newly found break frequencies are
broadly consistent with a disk truncated close to the light cylinder
with $\dot{M}\simeq10^{13}-5\times10^{14}$~g~s$^{-1}$ and a viscosity
parameter $\alpha \gtrsim$0.2.

\end{abstract}

\begin{keywords}
X-rays:
individual(IGR~J18245-2452, 2XMM~J102347.6+003841, XSS~J12270-4859) ---
stars: neutron --- X-rays: binaries --- globular clusters:
individual(M28) --- pulsars:
individual(PSR~J1824--2452I, PSR~J1023+0038, PSR~J1227-4853 ---
accretion, accretion discs)
\end{keywords}

\section{Introduction}
\label{sec:intro}

Millisecond radio pulsars (MSPs) have X-ray luminosities
L$_\mathrm{X}$ (0.5--10~keV) of order
10$^{29}$--10$^{33}$~erg~s$^{-1}$, powered by the loss of rotational
energy of a rapidly spinning low magnetic field neutron star
\citep{Possenti02}.
Low-mass X-ray binaries (LMXBs), when active or in outburst, show
typically L$_\mathrm{X}\sim$10$^{35}$--10$^{38}$~erg~s$^{-1}$ thanks
to the gravitational potential energy released during the accretion
process.
These two types of neutron star binaries represent two phases of an
evolutionary sequence, where LMXBs are thought to precede and produce
MSPs \citep{Alpar82}. 

The link between MSPs and LMXBs has been confirmed in the
last decade with the discovery of three neutron stars that alternate
between rotation-powered (MSP) and accretion-disk (LMXB) states, on
timescales of days to months \citep{Archibald09,Papitto13b,Bassa14}.
All three systems, known as ``transitional millisecond pulsars''
(tMSPs), have low-mass main-sequence companion stars in compact orbits
(period P$_\mathrm{orb}$$\simeq$4--11~hr).
Thus they belong to the growing class of compact binary or ``spider''
MSPs, in particular to the ``redback'' type \citep{Roberts13}.

\citet{Linares14c} studied the global behavior of redbacks and
identified three main X-ray states: i) (radio) {\it pulsar state}
(10$^{31}<$L$_\mathrm{X}<4\times$10$^{32}$~erg~s$^{-1}$); ii) {\it
  disk state}
(4$\times$10$^{32}$$<$L$_\mathrm{X}$$<$10$^{34}$~erg~s$^{-1}$) ; and
iii) {\it outburst state}
(10$^{34}$$<$L$_\mathrm{X}$$<$10$^{37}$~erg~s$^{-1}$)\footnote{Hereafter,
  all L$_\mathrm{X}$ are corrected for absorption and quoted in the
  0.5--10 keV band.}.
The outburst state, which shows thermonuclear X-ray bursts and
peculiar X-ray variability, has been seen only in
PSR~J1824--2452I/IGR~J18245--2452 \citep[M28-I hereafter, in the
  globular cluster M28; see][]{Papitto13b,Ferrigno14,Wijnands17}.
The pulsar state (PS) of most redbacks shows a characteristic
double-peaked orbital modulation of the X-ray flux
\citep{Bogdanov11b}, and no optical evidence for the presence of an
accretion disk.

The disk state, at the intersection between accretion and rotation
power, is of particular interest as it allows unprecedented studies of
underluminous accretion flows, pulsar winds and their mutual
interaction.
\citet{Linares14} discovered striking rapid X-ray variability in the
disk state of the transitional millisecond pulsar M28-I.
While in the disk state, M28-I switches rapidly between two clearly
distinct modes/states: the {\it disk-high mode} (DH; also known as
disk-active) with L$_\mathrm{X}$=3.9$\times$10$^{33}$~erg~s$^{-1}$ and
the {\it disk-low mode} (DL; also known as disk-passive) with
L$_\mathrm{X}$=5.6$\times$10$^{32}$~erg~s$^{-1}$.
\citet{Linares14c} pointed out that this phenomenon, labeled {\it
  X-ray mode switching}, is ubiquitous in the disk state of {\it all
  three} tMSPs \citep[see also][]{deMartino13,Tendulkar14}, yet it is
not observed in LMXBs at similar L$_\mathrm{X}$.
In fact, X-ray mode switching has been used since to identify new
candidate tMSPs \citep[e.g.,][]{Bogdanov15}.

\begin{table*}
\center
\footnotesize
\caption{X-ray observations of the disk and pulsar states of tMSPs analyzed in this work.}
\begin{minipage}{\textwidth}
\begin{tabular}{l c c c c c c c c c}
\hline\hline
System & Epoch & Mission  & $N_\mathrm{obs}$ & $T_\mathrm{res}$ & Exp.   & Long gtis  & FFTs  & $\nu_0$--$\nu_\mathrm{Nyq}$ & Total/BKG \\
(tMSP)  & (yr)  & (instr.) &                 & (ms)           & (ks)    & (ks)    & (nr.$\times \Delta T$,s)     & (Hz)                      & rate(c/s)  \\
\hline
\multirow{3}{*}{M28-I} & \multirow{2}{*}{2008} & \multirow{3}{*}{\thead[cc]{Chandra\\(ACIS-S)}}   & \multirow{2}{*}{2\footnote{Obs. IDs: 9132,9133}}
                       & \multirow{2}{*}{3141} & \multirow{2}{*}{199}   & DH: 101  & 3$\times$19299 & 5.2$\times$10$^{-5}$--0.053 & 0.08/1.8$\times$10$^{-4}$\\
                       &                       &     &  &                       &                        & DL: 67   & 9$\times$6433  & 1.6$\times$10$^{-4}$--0.16  & 0.014/1.8$\times$10$^{-4}$\\
                       &        2015           &     &  2\footnote{Obs. IDs: 16749,16750}  & 3241 & 60  & PS: 60   & 6$\times$9956  & 1.0$\times$10$^{-4}$--0.051 & 1.7$\times$10$^{-3}$/1.8$\times$10$^{-4}$\\
\hline
\multirow{3}{*}{\thead[cl]{PSR\\J1023+0038}} & 2013-         & XMM          & \multirow{2}{*}{18\footnote{Obs. IDs: 0720030101, 0742610101, 0748390-101/501/601/701, 0770581-001/101, 0783330301, 0784700201, 0794580-801/901, 0803620-201/301/401/501, 0823750-301/401}}
& \multirow{2}{*}{0.122} & \multirow{2}{*}{828}   & DH: 123  & 4$\times$4096 & 2.4$\times$10$^{-4}$--128 & 3.17/0.22\\
                       &          -2018          & (EPIC-pn)    &  &                       &                        & DL: 28   & 15$\times$768  & 1.3$\times$10$^{-3}$--683  & 0.44/0.07 \\
                                 &        2010             & Chandra    & 1\footnote{Obs. ID: 11075}  &      1141                 &           83           & PS: 83   & 4$\times$18695  & 5.3$\times$10$^{-5}$--0.22  & 0.038/1.9$\times$10$^{-4}$ \\
\hline
\multirow{3}{*}{\thead[cl]{XSS\\J12270-4859}} &  \multirow{2}{*}{2011}       & \multirow{2}{*}{XMM}          & \multirow{2}{*}{1\footnote{Obs. ID: 0656780901}}
& \multirow{2}{*}{0.122} & \multirow{2}{*}{30}   & DH: 7.2  & 2$\times$2048 & 4.9$\times$10$^{-4}$--256 & 2.98/0.12\\
                                 &                         &     &  &                       &                        & DL: 2.4   & 4$\times$512  & 2.0$\times$10$^{-3}$--1024  & 0.46/0.07 \\
                                 &        2014             & Chandra    & 1\footnote{Obs. ID: 16561}  &      3141                 &           30           & PS: 30   & 3$\times$9649  & 1.0$\times$10$^{-4}$--0.053 & 0.032/1.3$\times$10$^{-4}$ \\

\hline\hline
\end{tabular}
\end{minipage}
\label{table:obs}
\end{table*}

In this work we take a closer look at the aperiodic X-ray variability
of tMSPs in the disk and pulsar states, using {\it Chandra} and {\it
  XMM-Newton} observations of all three confirmed tMSPs: M28-I,
  PSR~J1023+0038 (J1023 hereafter) and XSS~J12270-4859 (J1227
  hereafter).
We find strong X-ray variability in the disk state (DH mode), in the
form of flat-top noise and a characteristic {\it break} frequency, all
typical of LMXBs in the hard state.
This gives X-ray evidence for the presence of an accretion disk in the
disk state of tMSPs, and the lowest break frequencies ever
detected in a LMXB.
We also discuss the implications of these new results on the
models proposed for the disk state of tMSPs.

\section{Analysis}
\label{sec:data}

We analyzed all {\it XMM-Newton} and {\it Chandra} observations of the three
confirmed tMSPs, available in mid 2021 and taken when they were in the
disk state.
For each observation we selected manually the ``long good time
intervals'' when the tMSPs were in the disk-active/high (DH) and
disk-passive/low (DL) states, using 10 and 100~s-resolution light
curves for {\it XMM-Newton} and {\it Chandra}, respectively.
We excluded intrinsic flaring states, a few background flaring periods
in the case of {\it XMM-Newton}, and those intervals shorter than about 1 and
0.3 ks for the DH and DL states, respectively (since we are interested
on long timescale variability).
This careful selection is necessary to study and quantify the X-ray
variability in the DH and DL states and results in just
$\sim$5--50\% of the data being used, depending on the source and
state.

\begin{figure*}
  \begin{center}
  \resizebox{0.69\columnwidth}{!}{\rotatebox{-90}{\includegraphics[]{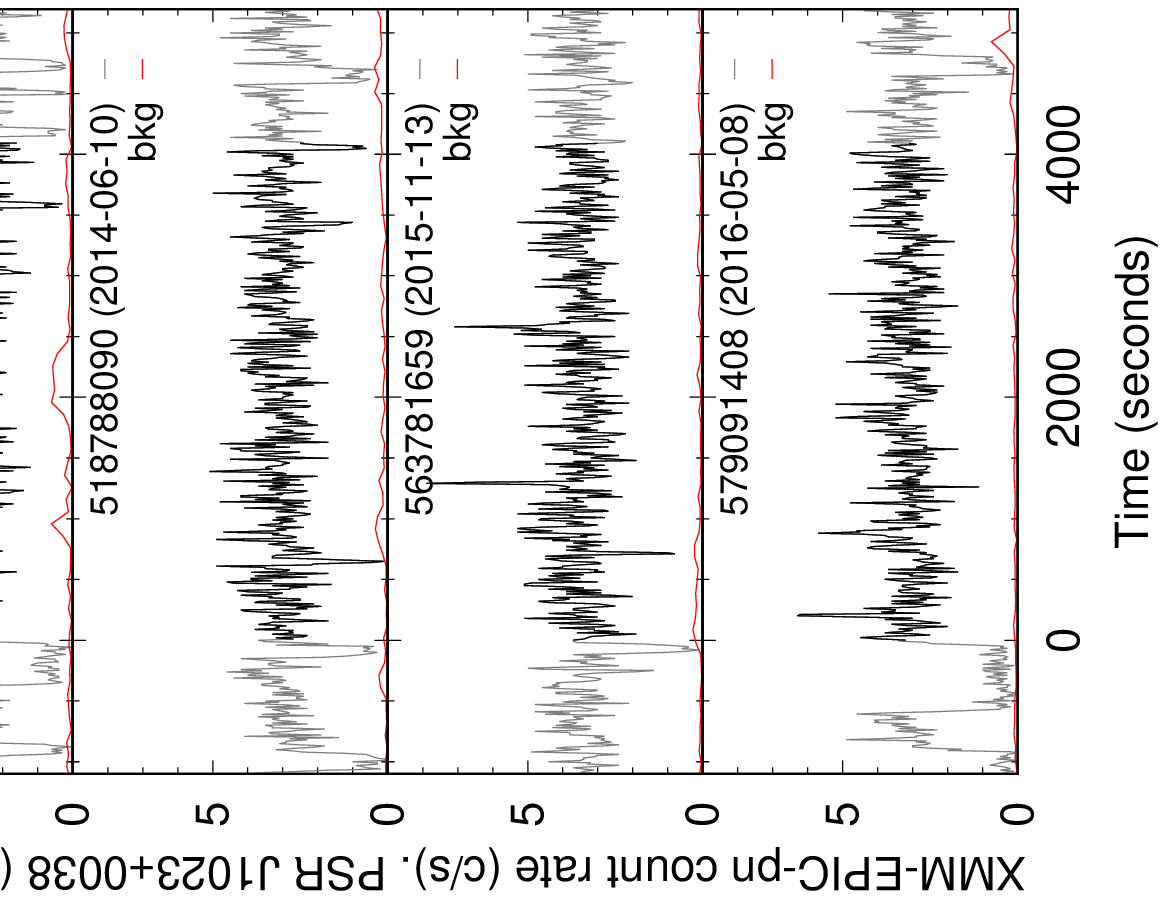}}}
  \resizebox{0.69\columnwidth}{!}{\rotatebox{-90}{\includegraphics[]{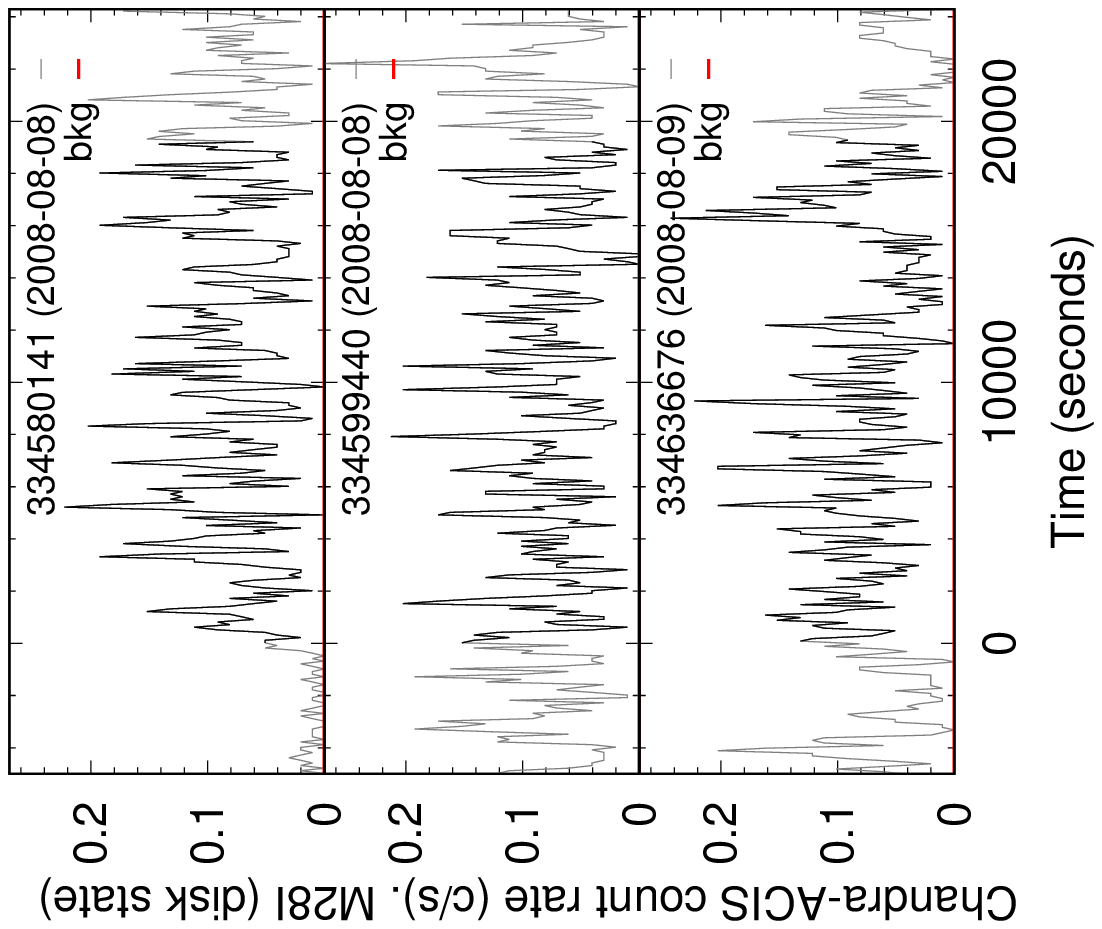}}}
  \resizebox{0.69\columnwidth}{!}{\rotatebox{-90}{\includegraphics[]{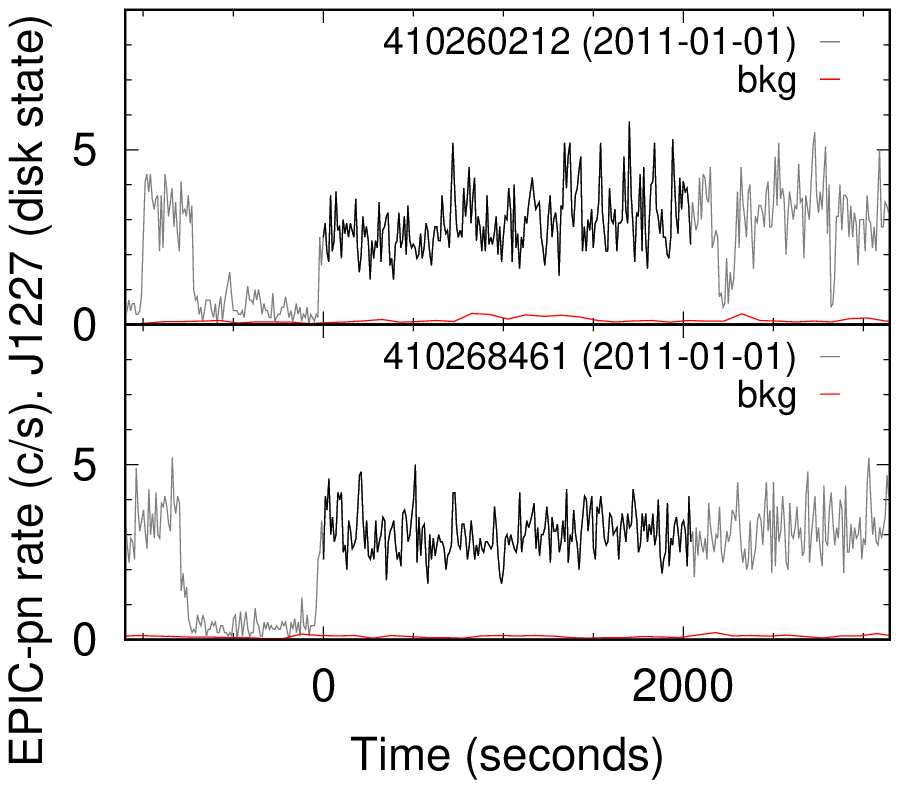}}}
  \caption{
X-ray light curves of the three tMSPs in the disk state, from left to
right: PSR~J1023+0038, M28-I and XSS~J12270-4859. Black solid lines
show the time intervals of the disk-high (DH) state used for aperiodic
timing analysis (FFTs). Gray lines show wider segments of the disk
state lightcurves, and red lines show background rates. The FFT start
times (in MET) and corresponding dates are indicated on each panel.
} %
    \label{fig:lcs}
 \end{center}
\end{figure*}

We also included selected {\it Chandra} observations of all three
systems in the pulsar state, in order to compare the X-ray
variability of these different states quantitatively.
Table~\ref{table:obs} shows the details of all analyzed observations,
and the resulting effective exposure times after summing all the long
good time intervals in the DH \& DL modes (``long gtis'').
We show in Figure~\ref{fig:lcs} light curves of the three tMSPs in
the disk state, as well as the exact time segments used to
extract all DH mode power spectra.

\subsection{XMM-Newton}
\label{sec:xmm}

We analyzed all EPIC-pn timing data of the disk state of the two tMSPs
J1023 and J1227, for a total exposure time of 828~ks and 30~ks,
respectively.
All {\it XMM-Newton}-EPIC-pn observations were reprocessed with {\sc EPPROC}
(SAS version 20180620-1732-17.0.0) to generate calibrated event lists.
Since the analysed {\it XMM-Newton} observations were taken in timing mode
(with a time resolution of 122~$\mu$s), the two-dimensional imaging
information is lost.
Filtered source event files were extracted with {\sc
  XMMSELECT/EVSELECT}, keeping events with energies in the 0.5--10~keV
range (and pattern 0--4), within a 7-pixel-wide box centered on the
source (RAWX 38).
These events were used to generate power spectra as explained below,
including only time intervals when the sources were in the DH and DL
state.
For J1023 and J1227 this led to average total
EPIC-pn count rates of about 3 and 0.5 c/s for the DH and DL states,
respectively (Table~\ref{table:obs}).

The background in {\it XMM-Newton} has multiple components, is energy
dependent and strongly variable (even outside proton flares).
We first monitored the proton flare background by extracting light
curves between 10--15~keV, in the same source extraction region, and
verified that no flares affect our DH/DL time intervals.
We also excluded the lowest (0.2--0.5~keV) energies and restricted our
analysis to the 0.5--10~keV band \citep[]{Burwitz04}, which reduces
the background rate by a factor $\sim$10.
To measure the background rate we extracted events in the same energy
range from one identical region far from the source extraction region
(centered on RAWX=10), where the count profile is approximately flat
and we expect a negligible source contribution.
This is not the case in brighter sources \citep[see, e.g., Appendix B
  in][]{Wijnands17}, but our tMSPs are faint so that the tail of the
PSF is fainter than the true background.
We also checked another region on the opposite side of the detector (at
RAWX=55), and obtained consistent background rates.
This gives our final measurement of the average background rate in the
same energy band (0.5--10~keV), the same region (7$\times$100~pixel
box) and the exact same time intervals used for the FFTs.
These background rates used for the power spectral rms normalization
are about 5\% and 15\% of the total rates for the DH and DL states,
respectively, as shown more precisely in Table~\ref{table:obs}.

\subsection{Chandra}
\label{sec:chandra}

We analyzed both {\it Chandra}-ACIS-S observations of M28 taken in
2008, using {\sc CIAO} (version 4.13).
These provide the only ACIS imaging data of the tMSP M28-I in the disk
state, for a total exposure time of 199~ks.
We extracted filtered source event files with {\sc DMCOPY}, keeping
events with energies in the 0.2--10~keV range, within a 3-pixel-radius
circle centered on M28-I \citep[source 23 in][]{Becker03}.
These events were used to generate power spectra as explained below,
including only time intervals when the source was in the DH and DL
state.
The resulting average total ACIS count rate was about 0.08 and 0.014
c/s for the DH and DL states, respectively (Table~\ref{table:obs}).

We also extracted power spectra of the pulsar state of J1023
and J1227 using {\it Chandra}-ACIS-S observations taken in
2010 and 2014, respectively (Table~\ref{table:obs}).
In this case we did not select specific time intervals, since the
orbital X-ray variability in the pulsar state is relatively smooth and
happens on timescales of several hours, longer than those we are
studying.
Since the background in {\it Chandra} is low and stable, and the
observations were taken in an imaging (TIMED/VFAINT) mode, we use a
nearby source-free region with the same radius to measure the average
background rate in each observation.
These background count rates used for the power spectral rms
normalization are about 0.2 and 1\% of the total rates for the DH and
DL states, respectively, as shown in Table~\ref{table:obs}.

\subsection{Timing analysis}
\label{sec:timing}

We binned the original event files (with time resolution
$T_\mathrm{res}$) by a factor $n$ to reach a Nyquist frequency
$\nu_\mathrm{Nyq}$=1/(2$n T_\mathrm{res}$).
We applied fast Fourier transforms (FFTs) on light curve segments with
duration $\Delta T$, obtaining power density spectra (PDS) of the DH
and DL states separately (with a minimum Fourier frequency
$\nu_0$=1/$\Delta T$; see Table~\ref{table:obs}).
The resulting PDS were averaged and rebinned logarithmically.
We explored the range $\Delta T$=[0.3--20]~ks, and chose the optimum
value in each case to reach the lowest possible Fourier frequencies.

A constant Poisson noise contribution was subtracted \citep[matching
  the level above 0.75$\times \nu_\mathrm{Nyq}$; see,
  e.g.,][]{Kleinwolt04}, and the PDS were normalized in the
rms-normalization using the average background rates
\citep[][see Section~\ref{sec:xmm}]{vanderklis95b}.
In this normalization, the integrated power, $I=\sum_{n=i}^{j} \nu_0
P_n = \overline{P} \Delta \nu$, gives the fractional variance, where
$\overline{P}$ is the average power between two frequencies $\nu_i$
and $\nu_j$ such that $\Delta \nu=\nu_j - \nu_i$.
We use this property to measure the fractional root-mean-squared
amplitude of the variability ($rms=\sqrt{I}$) from the
Poisson-subtracted, rms-normalized PDS in two frequency bands:
0.001--0.1~Hz (low frequency, LF) and 0.1--10~Hz (high-frequency, HF).
When calculating the uncertainties on the $rms$ we take into account a
$\sim$20\% systematic error from the {\it XMM-Newton} background rates used
in the PDS normalization (but in most cases the statistical error from
the dispersion of PDS powers dominates).
When the measured $I$ was consistent with zero in a given frequency
band, we placed 90\% upper limits on the $rms$ assuming a normal
distribution of powers, i.e., using 1.3 times the standard deviation
of powers in that band.

We fit the PDS of J1023 and M28-I in the DH state with a broken
power-law model with the slope/index before the break fixed at zero,
to account for the flat-topped noise that we find and report below
\citep[we chose this simple model to allow for direct comparison with the
  early work of][]{BH90}.
When necessary, we add one Lorentzian to account for additional power
at higher frequencies, as is customary in more recent studies of
accreting X-ray binaries \citep[e.g.,][]{Wijnands99}.

\begin{figure*}
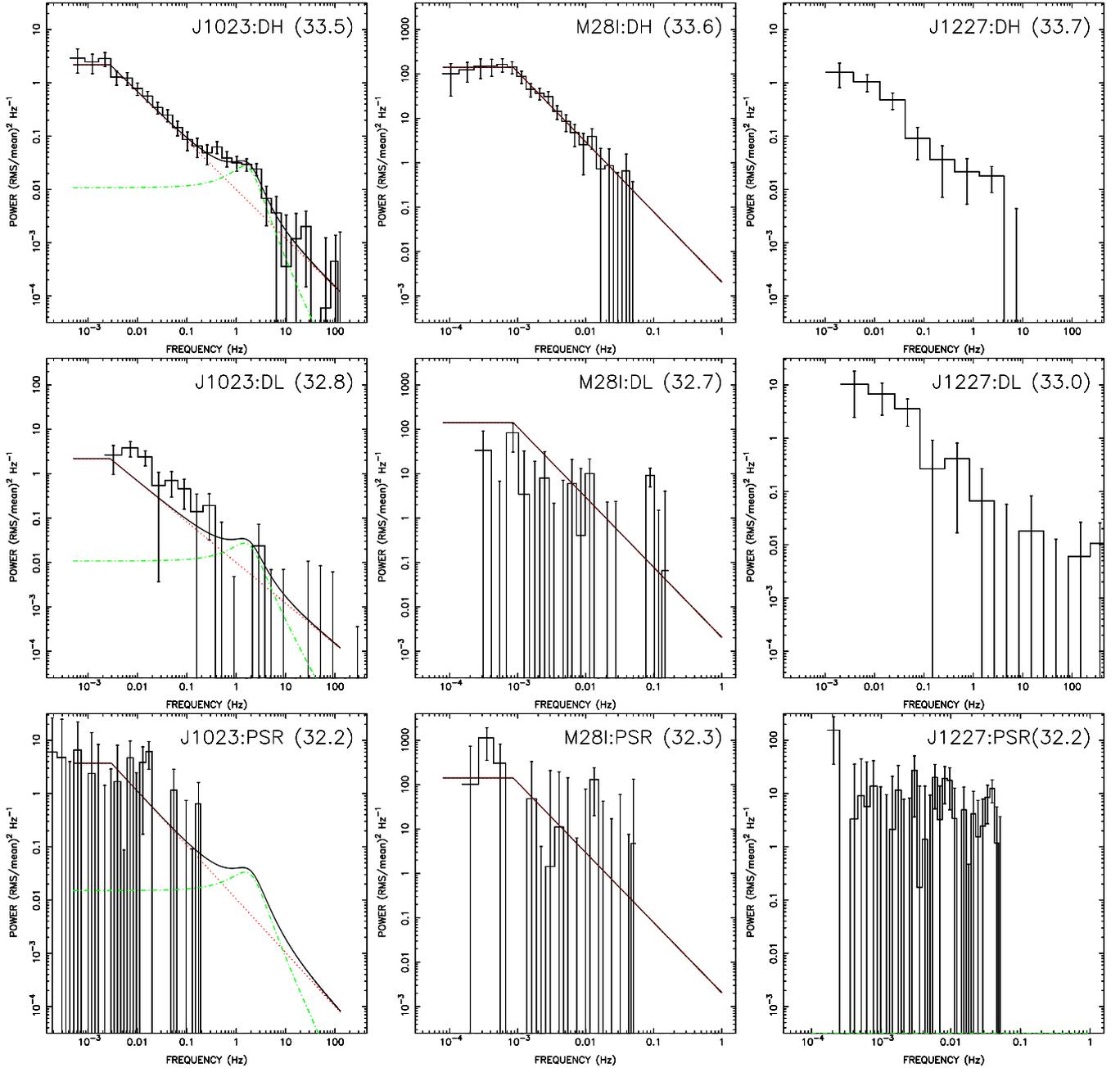

  \begin{center}
    \resizebox{0.69\columnwidth}{!}{\rotatebox{0}{\includegraphics[]{J1023_DH.ps}}}
    \resizebox{0.69\columnwidth}{!}{\rotatebox{0}{\includegraphics[]{M28I_DH.ps}}}
    \resizebox{0.69\columnwidth}{!}{\rotatebox{0}{\includegraphics[]{J12270_DH.ps}}}
    \resizebox{0.69\columnwidth}{!}{\rotatebox{0}{\includegraphics[]{J1023_DP.ps}}}
    \resizebox{0.69\columnwidth}{!}{\rotatebox{0}{\includegraphics[]{M28I_DP.ps}}}
    \resizebox{0.69\columnwidth}{!}{\rotatebox{0}{\includegraphics[]{J12270_DL.ps}}}
    \resizebox{0.69\columnwidth}{!}{\rotatebox{0}{\includegraphics[]{J1023_PSR.ps}}}
    \resizebox{0.69\columnwidth}{!}{\rotatebox{0}{\includegraphics[]{M28I_PSR.ps}}}
    \resizebox{0.69\columnwidth}{!}{\rotatebox{0}{\includegraphics[]{J12270_PSR.ps}}}
  \caption{
    {\it Top panels:} Power spectra of the three known tMSPs in the
    disk-high (DH) mode/state {\it (from left to right):} J1023, M28-I
    and J1227.
    {\it Middle panels:} Same for the disk-low (DL) mode/state.
    {\it Bottom panels:} Same for the pulsar (PSR) state
    (Section~\ref{sec:intro} for definitions).
    Lines show the broken power-law model fits to the corresponding DH
    power spectra (Section~\ref{sec:timing}).
    The logarithm of the 0.5--10~keV luminosity is shown between
    parentheses for each case (L$_\mathrm{X}$, in erg~s$^{-1}$).
} %
    \label{fig:pds}
 \end{center}
\end{figure*}

In order to verify that background variability does not affect our
results for the faintest tMSP states, we extracted PDS using events in
the {\it XMM-Newton} background region.
We find that the EPIC-pn (timing mode) background is strongly variable
at frequencies in the 0.1-100~Hz range ($>$1000\% fractional rms), but
it is not strongly variable at lower frequencies in our chosen time
intervals (fractional rms less than 51\% in the 0.0039--0.1~Hz
frequency band).
Since we select periods of low background rates for our {\it XMM-Newton}
analysis (and in general the background is only between 0.2\% and 15\%
of the total count rate, see Table~\ref{table:obs}), we can be
confident that the X-ray variability that we detect is intrinsic to
the tMSPs studied.

\section{Results}
\label{sec:results}

The disk state of tMSPs is strongly variable both within the DH and DL
modes, as can be already seen from the X-ray lightcurves shown in
Figure~\ref{fig:lcs}.
Beyond the X-ray mode switching discovered in 2013 \citep{Linares14},
we find intrinsic variability in the X-ray flux from tMSPs on a broad
range of frequencies, between 10$^{-4}$ and 10~Hz
(Figure~\ref{fig:pds}).
This corresponds to a wide range of timescales, between the orbital
(4.8--11~hr, i.e., $\sim$10$^{-5}$~Hz) and NS spin (1.7--3.9~ms, i.e.,
$\sim$10$^{2}$~Hz) periods.

\begin{table}
\center
\footnotesize
\caption{X-ray variability properties of the disk and pulsar states of tMSPs.}
\begin{minipage}{\textwidth}
\begin{tabular}{l c c c c}
\hline\hline
System & State & $\nu_\mathrm{b}$ &  rms$_\mathrm{LF}$ (\%) & rms$_\mathrm{HF}$ (\%) \\
(tMSP)  & /mode & (mHz)          & 0.001-0.1~Hz            & 0.1-10~Hz  \\
\hline
\multirow{3}{*}{M28-I}          & DH & 0.86$\pm$0.15 &  42$\pm$3 & - \\
                                & DL & -             &  $<$45      & - \\
                                & PS & -         & $<$167 & - \\
\hline
\multirow{3}{*}{J1023} & DH & 2.8$\pm$0.7\footnote{A double zero-centered Lorentzian model fit yields a higher\\ characteristic frequency, $\nu_\mathrm{max}$=12$\pm$3~mHz.}  & 18.4$\pm$1.0 & 32.1$\pm$2.9 \\
                                & DL & $<$8.5       & 35.1$\pm$5.2 & $<$49 \\
                                & PS & -         & $<$27 & - \\
\hline
\multirow{3}{*}{J1227}& DH & $<$23  & 17.4$\pm$1.7 & 27.0$\pm$6.7 \\
                                & DL & $<$37  & 60$\pm$8 & $<$70 \\
                                & PS & -         & $<$44 & - \\
\hline\hline
\end{tabular}
\end{minipage}
\label{table:res}
\end{table}

\subsection{Disk-high mode}
\label{sec:DH}

We find that the disk-high state (DH mode) of J1023 shows
flat-topped broad-band noise with a break frequency
$\nu_\mathrm{b}$=2.8$\pm$0.7~mHz (see Figure~\ref{fig:pds} and
Table~\ref{table:res}).
The variability is strong, with a fractional rms amplitude
$rms$=37.0$\pm$2.6~\% in the 0.001--10~Hz frequency band
(Section~\ref{sec:timing} for details).
We obtain a good fit to the DH power spectrum
($\chi^2$/d.o.f.=35.0/36) with the simple broken power-law plus
Lorentzian model shown in Figure~\ref{fig:pds} (top left).
For this Lorentzian component, we find a centroid frequency
$\nu_\mathrm{c}$=1.5$\pm$0.5~Hz, a full-width at half-maximum of
2.4$\pm$1.1~Hz and $rms=32\pm7 \%$.
This component thus peaks around 1~Hz, with a quality factor
$Q=\nu_\mathrm{c}/FWHM \simeq 0.6$.
The variability at lower frequencies (0.001--0.1~Hz) is less strong,
with $rms_{LF}$=18.4$\pm$1.0~\% (Table~\ref{table:res}).
The power-law index above $\nu_\mathrm{b}$ is consistent with 1
(0.92$\pm$0.07).
Interestingly, upon inspection of the individual PDS from the DH state
of J1023, we also find that both $\nu_\mathrm{b}$ and the integrated
power are approximately constant between 2013 and 2016.

Furthermore, we find that the tMSP M28-I in the disk state (DH mode;
Fig.~\ref{fig:pds}, top center) shows very similar and even stronger
flat-topped noise, with a lower break frequency
$\nu_\mathrm{b}$=0.86$\pm$0.15~mHz and $rms_\mathrm{LF}$=42$\pm$3~\%
(Table~\ref{table:res}).
The power-law index above the break is slightly higher too
(1.6$\pm$0.2).
In this case we cannot study variability at higher frequencies
($\gtrsim$0.1~Hz) since we are limited by the time
resolution (Table~\ref{table:obs}).

For the tMSP J1227 the X-ray variability is less
constrained, as there is only one 30-ks {\it XMM-Newton} observation of the
disk state in timing mode (much less than the 828 ks available for
J1023).
Still, we find strong variability between 10$^{-3}$ and 1~Hz
(Figure~\ref{fig:pds}, top right).
The fractional rms amplitude in the 0.001--10~Hz frequency band is
$rms$=32$\pm$6~\%.
Like in the DH mode of J1023, this variability is stronger at
higher frequencies (cf. $rms_\mathrm{HF}$=27.0$\pm$6.7~\% and
$rms_\mathrm{LF}$=17.4$\pm$1.7~\%; Table~\ref{table:res}).
Even though the DH mode power spectrum starts at
$\nu_0$=2.0$\times$10$^{-3}$~Hz, we can use it to constrain the break
frequency and we estimate a 90\% upper limit of $\nu_\mathrm{b}
<$ 23~mHz (a broken power-law fit yields break frequencies consistent
with 0 at the 2.4$\sigma$ level).
Thus, in this case we cannot measure $\nu_\mathrm{b}$ but the upper
limits are consistent with the values we find for the two other tMSPs.

\subsection{Disk-low mode}
\label{sec:DL}

The disk-low (DL, or ``passive'') state of tMSPs shows different X-ray
variability, as can be seen from Figure~\ref{fig:pds} (middle panels).
We do not detect a low-frequency break to a flat-topped noise, down to
the lowest available frequencies (between 0.2 and 2~mHz;
Table~\ref{table:obs}).
We set upper limits on $\nu_\mathrm{b}$ of $<$8.5~mHz and $<$37~mHz for
the DL mode of J1023 and J1227, respectively
(broken power law fits give a $\nu_\mathrm{b}$ consistent with 0
within 1--2.5 sigma, which we use to estimate 90\% upper limits).
These are consistent with the break frequencies that we find in the DH
mode, so we cannot assess with the current data if the flat-topped
noise persists in the DL mode.

\begin{figure*}
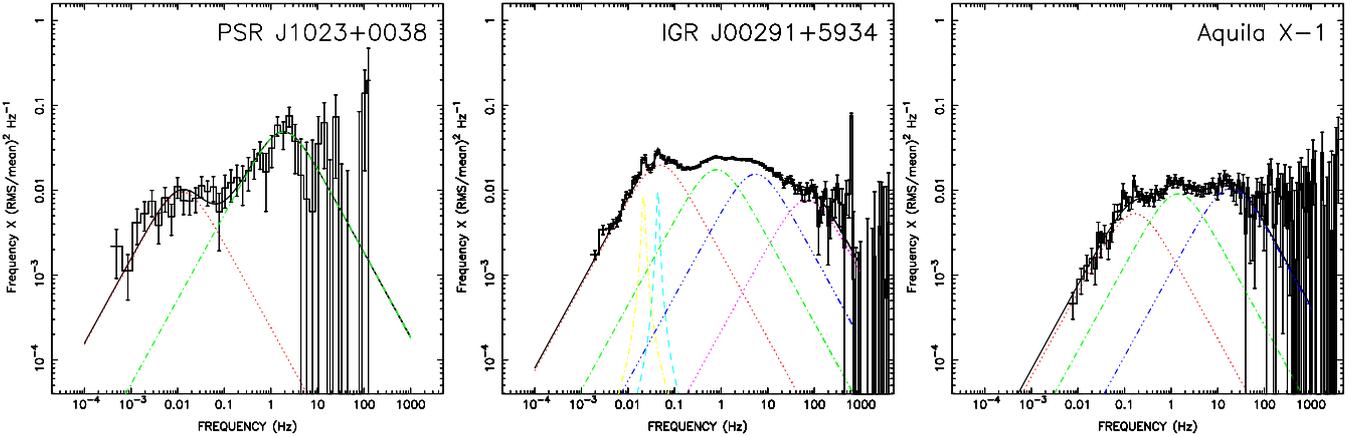

  \begin{center}

    \resizebox{0.69\columnwidth}{!}{\rotatebox{0}{\includegraphics[]{J1023_2Lor_PFvF_DH.ps}}}
    \resizebox{0.69\columnwidth}{!}{\rotatebox{0}{\includegraphics[]{IGRJ00291_6Lor_PFvF_HS.ps}}}
    \resizebox{0.69\columnwidth}{!}{\rotatebox{0}{\includegraphics[]{AqlX-1_3Lor_PFvF_HS.ps}}}

  \caption{
Power spectra (power$\times$frequency vs. frequency) of the tMSP J1023
in the DH state {\it(left)} and the NS LMXBs IGR~J00291+5934
{\it(center)} and Aquila~X-1 {\it(right)} in the hard state
\citep{Linares09d}. Multi-Lorentzian best-fit functions are shown with lines for
each case, where the lowest-frequency component (red dashed lines)
represents the break in the power spectra.
} %
    \label{fig:LMXB}
 \end{center}
\end{figure*}

Interestingly, we find that the low-frequency variability is stronger
in the DL mode, going from rms$_\mathrm{LF}$=18\% to
rms$_\mathrm{LF}$=35\% in the case of J1023.
The same is true for J1227, within the available constraints
(Table~\ref{table:res}).
On the other hand, the strong rapid variability around 1~Hz in the DH
mode is not detected in the DL mode (as can be seen from the power
spectra in Figure~\ref{fig:pds}).
However, the upper limits on the amplitude of the DL mode
high-frequency variability (rms$_\mathrm{HF} <$49 \% in J1023) are
consistent with the values of the DH mode.
The constraints on the X-ray variability of the DL mode of M28-I are
weaker, since the source is further away and fainter than the two
other tMSPs.
To summarize, low-frequency variability is stronger when going from
the more luminous DH mode to the less luminous DL mode.

\subsection{Pulsar state}
\label{sec:PS}

For completeness, we also quantify the aperiodic X-ray variability of
the pulsar state (PS), using the observations and PDS listed in
Table~\ref{table:obs}.
With an L$_\mathrm{X}$ more than ten times lower, and exposure times
typically shorter than those accumulated in the DH state, the
constraints are weaker.
We do not detect intrinsic source variability in the PS state on
timescales between $\sim$10--1000 s (the power in the
10$^{-3}$-10$^{-1}$ Hz band is consistent with that expected from
Poissonian counting noise).
We set upper limits of $rms_\mathrm{LF} <$27\%, 44\% and 167\% for
J1023, J1227 and M28-I, respectively.
Thus for both J1023 and J1227, we find that the fractional rms
amplitude in the pulsar state is significantly lower than that in the
disk state.

\section{Discussion}
\label{sec:discussion}

We have presented the discovery of intrinsic broad-band X-ray
variability in the DH state-mode of the tMSPs J1023 and M28-I, in the
form of flat-topped noise which breaks (to an approximately 1/$\nu$
form) above $\nu_\mathrm{b}\simeq$2.8 and 0.86~mHz, respectively
(Section~\ref{sec:DH}).
This provides a new observed signature of the presence of an accretion
disk in the X-ray emission of the disk state of tMSPs, since
flat-topped noise is regularly seen in disk-accreting LMXBs in their
hard states \citep[e.g.,][]{vanderKlis06,Linares09d}.
Our results also open a new way to potentially identify tMSP
candidates in the disk state, based on their aperiodic X-ray
variability.
This is especially important in Globular clusters, where identifying
the presence of an accretion disk via optical spectroscopy is
challenging \citep[but see][]{Cohn13}.
Previous X-ray timing studies of tMSPs did not detect this flat-topped
noise, perhaps due to the careful time interval selection required, or
because those have focused on lower (orbital) or higher (spin)
variability frequencies \citep[e.g.,][]{Tendulkar14,CotiZelati19}.

The X-ray luminosity of J1023 and M28-I in the DH state-mode is
L$_\mathrm{X}$=2.9 and 3.9$\times$10$^{33}$~erg~s$^{-1}$,
respectively \citep{Linares14c}.
This corresponds to 1--2$\times$10$^{-5}$ L$_\mathrm{Edd,NS}$ (where
L$_\mathrm{Edd,NS}$=2.3$\times$10$^{38}$~erg~s$^{-1}$ is the Eddington
luminosity of a 1.8~M$_\odot$ NS), i.e., a highly sub-Eddington
low-level accretion regime.
As noted previously, the L$_\mathrm{X}$ of the disk state of tMSPs is
remarkably constant on timescales of years (mode switching aside),
while we have found $\nu_\mathrm{b}$ to be also constant in J1023
within the errors (Section~\ref{sec:DH}).
This shows that the disk fluctuates on short (sub-second to hours)
timescales but is remarkably stable on long (year) timescales.

We can also compare the disk state L$_\mathrm{X}$ with the spin-down
luminosity which powers pulsar winds, L$_\mathrm{SD} \simeq $ 4.3 and
9.0$\times$10$^{34}$~erg~s$^{-1}$ for J1023 and J1227, respectively
\citep{Deller12,Roy15}.
We see that in the DH state L$_\mathrm{X}$ is 5--7\% of
L$_\mathrm{SD}$ (this fraction increases during flares, or if we take
into account bolometric corrections).
The DL state of J1023 and J1227 has L$_\mathrm{X}$=5.7 and
9.1$\times$10$^{32}$~erg~s$^{-1}$, respectively, which is 1\% of
L$_\mathrm{SD}$.
If we assume that accretion takes place down to the light cylinder
radius R$_\mathrm{lc} = c/(2 \pi \nu_\mathrm{spin})$ (81 and 188 km
for J1023/J1227 and M28-I, respectively), the accretion luminosity
released L$_{accr} = GM\dot{M}/R_{lc}$ becomes similar to the observed
L$_\mathrm{X} \sim 10^{33}-10^{34}$~erg~s$^{-1}$ for $\dot{M} \sim
10^{13}-10^{14}$~g~s$^{-1}$.
Both energy sources are plausible, and as pointed out by previous work
the disk state of tMSPs is powered most likely by a mix of NS rotation
and accretion power \citep[e.g.,][]{Linares14c,Campana16}.
This allows us to study the interaction between NS magnetospheres,
pulsar winds and underluminous accretion flows, as we discuss in more
detail below.

\subsection{Low-level accretion: comparison with LMXBs}
\label{sec:LMXB}

Our results reveal, to our knowledge, the lowest characteristic
(break) frequencies ever detected in the X-ray variability of LMXBs.
\citet{Tomsick04} found $\nu_\mathrm{b}\simeq$3.5~mHz in a black hole
(BH) LMXB accreting at
L$_\mathrm{X}$=3.3$\times$10$^{34}$~erg~s$^{-1}$=2.5$\times$10$^{-5}$
L$_\mathrm{Edd,BH}$, where we take
L$_\mathrm{Edd,BH}$=1.3$\times$10$^{39}$~erg~s$^{-1}$ as the Eddington
luminosity of a 10 M$_\odot$ BH \citep[we have converted their X-ray
  flux to the 0.5-10~keV band and used an updated distance of 4.5~kpc,
  from][]{Corral16}.
In a study of very hard NS-LMXBs, \citet{Wijnands17} found PDS breaks
down to $\sim$10$^{-2}$~Hz.
We show in Figure~\ref{fig:LMXB} the power spectrum of J1023 in the DH
state-mode (now in power$\times$frequency form), compared to those of
the accreting millisecond X-ray pulsars IGR~J00291+5934 and
Aquila~X-1.
The lower frequencies in J1023 are readily apparent: $\nu_\mathrm{b}$
is 1--3 orders of magnitude lower than what is seen in most NS LMXBs
in the hard state \citep{Linares09d}.
It is also interesting to note that J1023 and J1227 in the DH state
show strong variability around 1~Hz ($rms_\mathrm{HF}$ about 30\%,
Table~\ref{table:res}), stronger than what we find at lower
frequencies ($rms_\mathrm{LF}$ about 18\%).
This is different than LMXBs, where the fractional variability in
those frequency ranges is similar, and could perhaps reflect
additional variability imprinted on the inner accretion disk by the
interaction with the pulsar wind.

\begin{figure}
  \begin{center}
  \resizebox{1.0\columnwidth}{!}{\rotatebox{-90}{\includegraphics[]{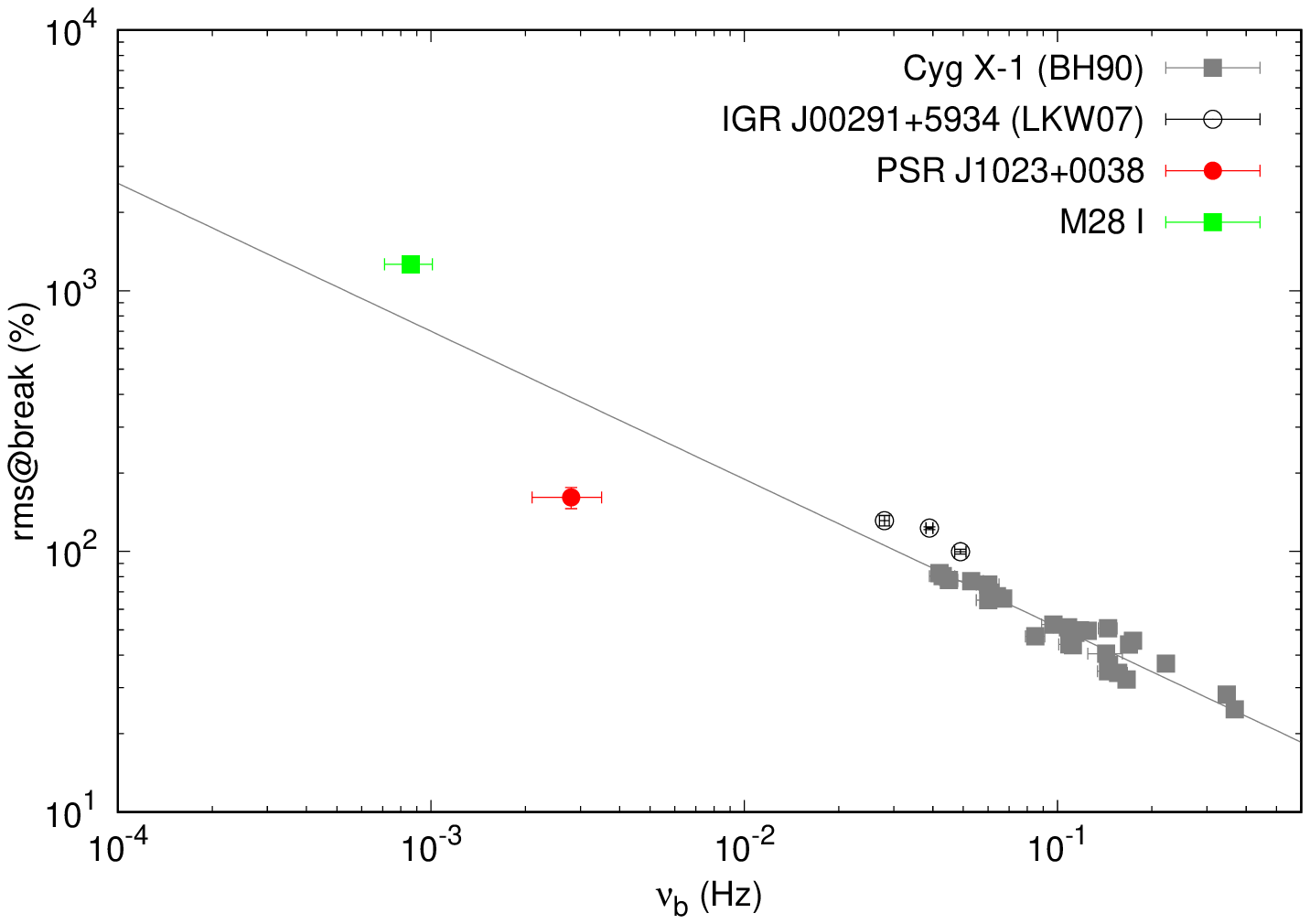}}}
  \caption{
Anti-correlation between the flat top power level and the break
frequency for the hard state of Cygnus~X-1 \citep{BH90}, the hard
(``extreme island'') state of IGR~J00291+5934 \citep{Linares07} and
the DH state of the two tMSPs analyzed herein.
} %
    \label{fig:BH}
 \end{center}
\end{figure}

It is widely accepted that the X-ray variability in LMXBs is generated
in the accretion flow, so that most variability frequencies reflect
dynamical and/or viscous timescales in the disk.
An interesting property of the flat-topped noise in the hard state of
disk-accreting X-ray binaries was found by \citet{BH90} from EXOSAT
studies of the black hole system Cygnus~X-1: the power level of the
flat-top and the break frequency $\nu_\mathrm{b}$ are anti-correlated.
We measure this power level in the DH state of J1023 and M28-I, by
averaging the PDS bins below $\nu_\mathrm{b}$ in each case, and
convert it to fractional rms in order to compare with previous
results.
As show in Figure~\ref{fig:BH}, the tMSPs J1023 and M28-I are broadly
consistent with the anti-correlation established for X-ray binaries at
higher luminosities, and extend it by 2 orders of magnitude in
$\nu_\mathrm{b}$, down to the mHz range.
This strongly suggests that the flat-topped noise in tMSPs is produced
by the same physical mechanism that produces this component in X-ray
binaries.

\begin{figure}
  \begin{center}
  \resizebox{1.0\columnwidth}{!}{\rotatebox{-90}{\includegraphics[]{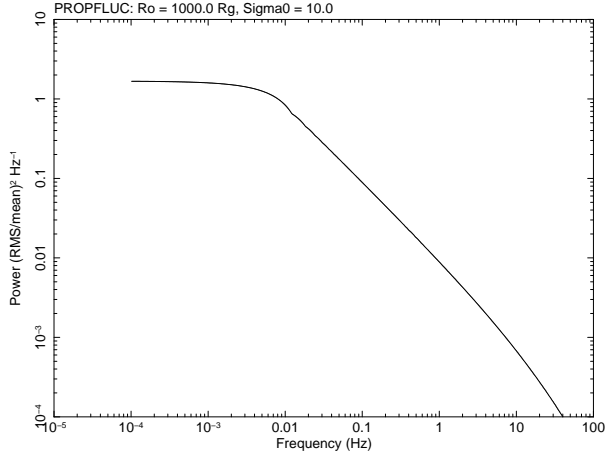}}}
  \caption{
Simulated X-ray power spectrum using the propagating fluctuations
model \textsc{propfluc} \citep{Ingram12} and an inner disk radius
(outer radius of the inner hot flow) set at 1000~R$_\mathrm{g}$.
} %
    \label{fig:propfluc}
 \end{center}
\end{figure}

The flat-topped noise in the hard state of LMXBs is thought to be
caused by fluctuations in the mass accretion rate at a range of radii
in the accretion disk/flow, which propagate inwards and are superposed
in the innermost regions giving rise to variability in L$_\mathrm{X}$
at a range of frequencies \citep{Lyubarskii97,Psaltis00,Done07}.
In most models the break frequency is given by the viscous timescale
of the inner edge of the disk, at a radial distance R$_\mathrm{in}$
from the center of the accreting NS or BH.
The power-spectral model \textsc{propfluc} \citep{Ingram12,Ingram13}
implements this by assuming that accretion rate fluctuations propagate
through an inner (precessing) hot flow, located within the truncated
disk inside R$_\mathrm{in}$, where the surface density radial profile
is given by a smoothly broken power law.
We attempted to match the PDS from J1023 in the DH state by varying
the most important parameters of \textsc{propfluc}: the outer radius
of the hot flow R$_\mathrm{o}$ (which in this model is the inner
truncation radius of the disk), and the normalization of the surface
density profile in the flow ($\Sigma_0$).
We set the compact object mass to 1.8~M$_\odot$, the bending wave
radius to 4.6~R$_\mathrm{g}$ (where $R_\mathrm{g}=GM/c^2$=2.7~km is
the gravitational radius) and all QPO normalizations to 0, while
fixing the fractional variability per decade in radius at Fvar=0.2.
The remaining parameters are fixed at their default values and have
little or no impact on $\nu_\mathrm{b}$ \citep[$\kappa$=3,
  $\lambda$=0.9, $\zeta$=0; see][for details]{Ingram12}.
Figure~\ref{fig:propfluc} shows that \textsc{propfluc} can roughly
reproduce the shape, power and order of magnitude of $\nu_\mathrm{b}$
in the PDS of J1023 in the DH state, for large values of
R$_\mathrm{o}\sim$1000~R$_\mathrm{g}$=2.7$\times$10$^8$~cm.
This apparently large truncation radius cannot be easily interpreted,
since \textsc{propfluc} models the hot inner flow of BH X-ray
binaries and it uses viscous frequencies which are higher than those
of a standard accretion disk (by 2-4 orders of magnitude in the
innermost regions; see Fig.~\ref{fig:freqs}).

\subsection{Mode switching in tMSPs: where is the disk truncated?}
\label{sec:mode}

One of the most interesting questions brought by the discovery of
tMSPs is how and where the disk is truncated in their
(intermediate/sub-luminous) disk state.
\citet{Linares14} proposed in their ``tug-of-war'' model that the disk
is truncated just outside the light cylinder in the DL state, and
fluctuations in $\dot{M}$ move it inside R$_\mathrm{lc}$ in the
transition to the DH state/mode.
They noted that in this case the rotation-powered pulsar would be
active at least in the DL mode.
This scenario seemed to be supported by \citet{Campana16}, who found
X-ray spectral evidence for an accretion disk truncated at R$_{in}
\simeq 20$~km in the DH state, while constraining R$_{in} \gtrsim
200$~km in the DL state.
Some of the early models for X-ray mode switching and the disk state
of tMSPs invoke a disk with a high $\dot{M} \sim 10^{15}$~g~s$^{-1}$
truncated close to the co-rotation radius \citep[R$_\mathrm{co}$=26~km
  for J1023/J1227, where $\nu_\mathrm{spin}$ equals the Keplerian
  frequency $\nu_\mathrm{k}$;][]{Papitto14}.

However, the discovery of X-ray \citep{Archibald15,Papitto15} and
optical \citep{Ambrosino17} pulsations in the DH state changed this
picture.
\citet{Papitto19} argued that the optical pulsations cannot be
produced by magnetically channeled accretion on the NS polar cap, and
also suggested that a rotation-powered pulsar is active in the disk
state.
There have also been claims of a disk truncated further out, at
200-300~R$_\mathrm{lc}$, based on models of the observed (broadband)
spectral energy distribution \citep{Coti14b,HernandezSantisteban16}.

\begin{figure}
  \begin{center}
  \resizebox{1.0\columnwidth}{!}{\rotatebox{-90}{\includegraphics[]{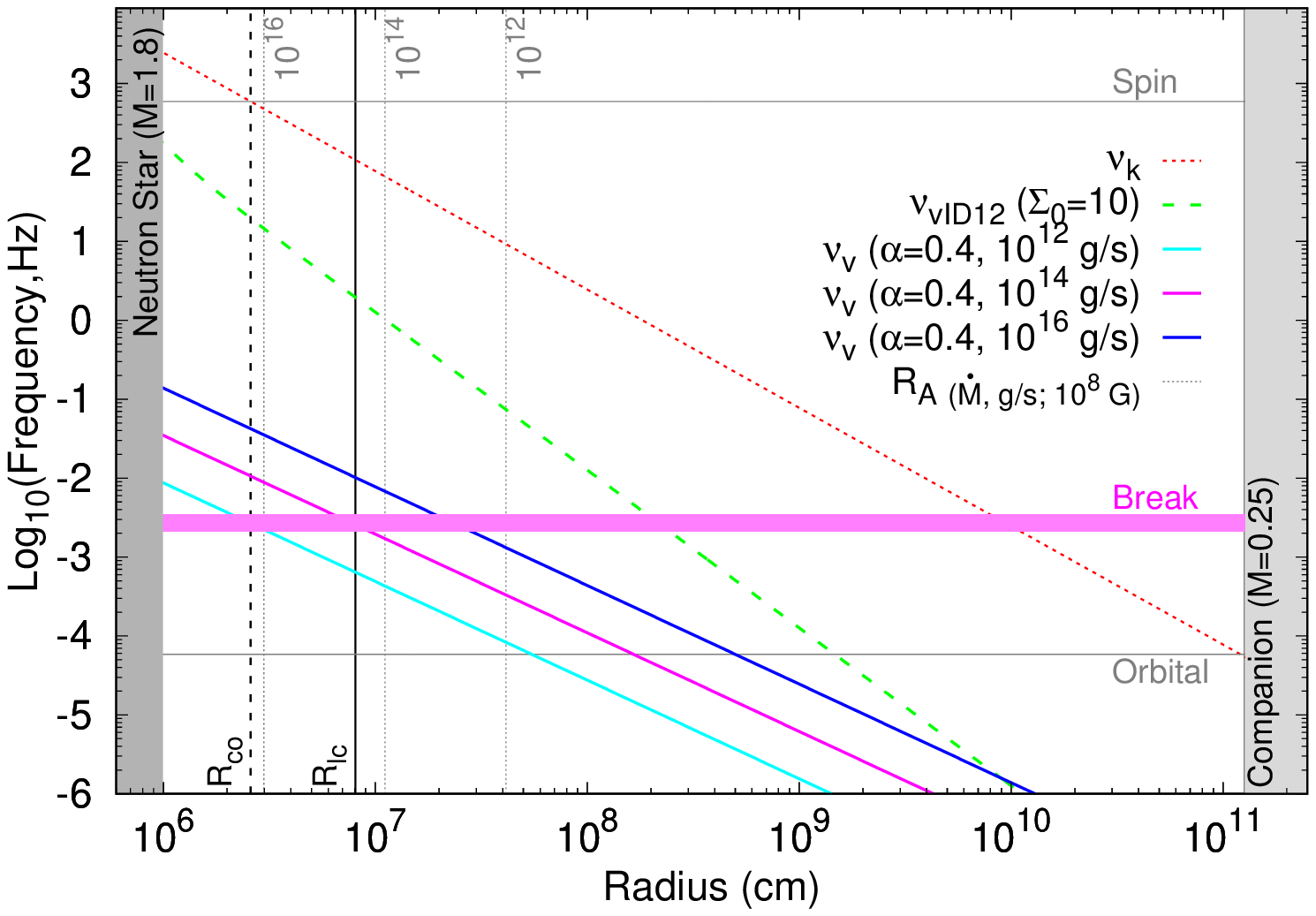}}}
  \caption{
    Frequency-radius plane for the tMSP J1023, showing viscous
    frequencies for a standard ``alpha'' disk at different mass
    accretion rates (cyan/magenta/blue solid lines, from
    \citealt{Frank02}) and for a hot inner flow (green dashed line,
    from \citealt{Ingram12}). We also show the Keplerian frequency
    assuming a NS mass of 1.8 M$_\odot$ (red dashed line) and the
    orbital separation for a 0.25 M$_\odot$ companion star.  The spin
    and orbital frequencies are shown with horizontal lines.  The
    magenta horizontal stripe shows our measurement of
    $\nu_\mathrm{b}$=2.8 mHz for the DH state of J1023.  The
    co-rotation and light-cylinder radii are shown with black dashed
    and solid lines, respectively. Gray dotted vertical lines show the
    location of the Alfven radius for different mass accretion rates,
    as indicated.
} %
    \label{fig:freqs}
 \end{center}
\end{figure}

More recently, \citet{Veledina19} presented a detailed physical model
of the disk state of tMSPs, where they propose that the disk is
truncated outside R$_\mathrm{lc}$ in the DH state, and enters the
light cylinder in the transition to the DL state.
Note this is opposite to the original ``tug-of-war'' model, which
proposed an inside-out transition to the less luminous DL mode.
We can use our results to test these models, under the common
assumption that the break frequency $\nu_\mathrm{b}$ is the viscous
frequency at the inner edge of the disk (see
discussion in Section~\ref{sec:LMXB}).
We show in Figure~\ref{fig:freqs} the viscous frequency
$\nu_\mathrm{v} \propto R^{-5/4}$ for a standard ``alpha'' disk
\citep{Shakura73} with viscosity parameter $\alpha=0.4$ and different
values of $\dot{M}=10^{12},10^{14},10^{16}$~g~s$^{-1}$ \citep[from
  Equation 5.69 in][]{Frank02}.
This is compared to the relevant frequencies for J1023:
$\nu_\mathrm{b}$ and $\nu_\mathrm{k}$ \citep[assuming a 1.8~M$_\odot$
  NS,][]{Shahbaz19}.
If the DH$\rightarrow$DL mode transition involves a decrease in
R$_\mathrm{in}$ by a factor 3 (say from 150~km to 50~km across
R$_\mathrm{lc}$, with constant average $\dot{M}$ and $\alpha$), then
we would expect $\nu_\mathrm{b}$ to increase by a factor
$3^{5/4}\simeq 3.9$ due to the higher viscous frequencies at smaller
radii in the disk (reaching about 11~mHz for the DL state of J1023).
We have placed an upper limit on $\nu_\mathrm{b} < 8.5$~mHz in the DL
state (Table~\ref{table:res}), which seems to disfavour an outside-in
DH$\rightarrow$DL transition of this magnitude.
Furthermore, our discovery of X-ray variability typical of LMXBs
favors an accretion-powered DH state, instead of a mostly
rotation-powered one.

\begin{figure}
  \begin{center}
  \resizebox{1.0\columnwidth}{!}{\rotatebox{-90}{\includegraphics[]{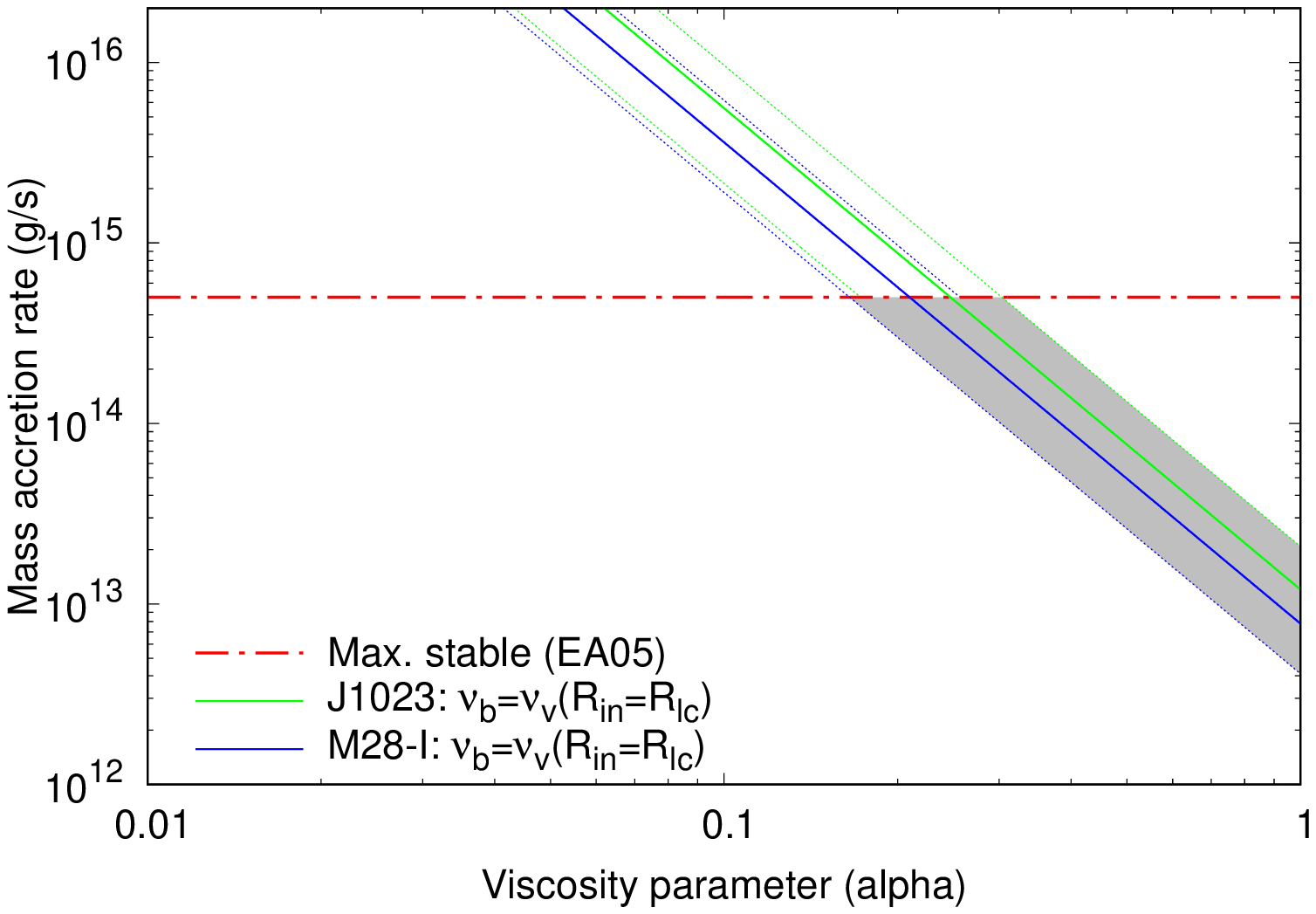}}}
  \caption{
Constraints on $\dot{M}$ and $\alpha$ (gray-shaded region) from our measured
$\nu_\mathrm{b}$ in J1023 and M28-I, assuming that $\nu_\mathrm{b} =
\nu_\mathrm{v}(R_\mathrm{in}=R_\mathrm{lc})$. The horizontal dot-dashed
line shows the maximum stable $\dot{M}$, from \citet{Eksi05}. Dotted lines show the projected uncertainties on $\nu_\mathrm{b}$.} %
    \label{fig:mdota}
 \end{center}
\end{figure}

If we neglect other forces acting on the innermost accretion flow,
magnetic pressure will disrupt the flow inside the Alfven radius
(R$_\mathrm{A}$, for spherical accretion, see e.g. \citealt{Frank02}).
In our case R$_\mathrm{A}$ is between 30 and 400~km, as shown in
Figure~\ref{fig:freqs}, for a 10$^8$~G surface magnetic field strength
and the same values of $\dot{M}=10^{12},10^{14},10^{16}$~g~s$^{-1}$
covering the most likely range for the disk state of tMSPs.
One should keep in mind that this is a simple prescription, which
ignores the potentially important effect of the pulsar wind on the
accretion disk.
\citet{Eksi05} studied such effect analytically and found that there
are stable disk solutions around an active rotation-powered pulsar,
but these are restricted to a narrow range of R$_\mathrm{in}\simeq
1-2$~R$_\mathrm{lc}$.
This may be the physical reason behind the long-term stability of the
disk state of tMSPs: the disk can exist only with a narrow range of
inner radii.
Furthermore, they find that such stable disk solutions can exist only
for $\dot{M}\simeq10^{12}-5\times10^{14}$~g~s$^{-1}$: at higher rates
the disk enters R$_\mathrm{lc}$, and at lower rates it is blown away
by the pulsar wind.

We can see from Figure~\ref{fig:freqs} that, if we assume that the
disk is truncated near the light cylinder (R$_\mathrm{in}\simeq
$R$_\mathrm{lc}$), the observed break frequency of J1023
($\nu_\mathrm{b}$=2.8~mHz) matches the viscous frequency of a disk
with $\alpha$=0.4 and $\dot{M}=10^{14}$~g~s$^{-1}$.
Interestingly, the same is true for the break frequency in M28-I:
$\nu_\mathrm{b}$=0.86~mHz is lower but the light cylinder is larger
(R$_\mathrm{lc}$=188~km), so the viscous frequency there is lower.
This offers a possible explanation for the significantly different
values of $\nu_\mathrm{b}$ that we find.

If we now put our main results for J1023 and M28-I together, assuming
that in both cases $\nu_\mathrm{b}$ corresponds to $\nu_\mathrm{v}$ at
R$_\mathrm{in}$, we can place constraints on the disk parameters
$\dot{M}$ and $\alpha$ (as shown in Figure~\ref{fig:mdota}).
To summarize, our results are broadly consistent with a disk truncated
at R$_\mathrm{in}\simeq $R$_\mathrm{lc}$ with
$\dot{M}\simeq5\times10^{12}-5\times10^{14}$~g~s$^{-1}$ \citep[where the upper
limit comes from][]{Eksi05} and a viscosity parameter $\alpha
  \gtrsim$0.2.
This range is consistent with most observational constraints on the
viscosity parameter $\alpha$ of a thin and fully ionized disk, derived
from studies of accreting white dwarfs and black holes \citep[see,
  e.g., ][and references therein]{King07,Tetarenko18}.

Without regard to the exact location of R$_\mathrm{in}$, our work
reveals strong fluctuations in $\dot{M}$ that may trigger the
transitions between X-ray modes that are seen in the disk state.
Future studies may be able to find a flat-topped noise in the DL
state/mode as well, perhaps with a different break frequency, and may
even allow a sensitive search for quasi-periodic oscillations.
To illustrate this, we estimate that {\it Athena} (X-IFU) will be able
to collect 0.5-10~keV count rates from J1023 of about 25, 5 and
0.7~c~s$^{-1}$ in the DH, DL, and PS states, respectively (between 8
and 20 times those listed in Table~\ref{table:obs}).

\section{Acknowledgments}

We thank A. Ingram for insightful discussions on the \textsc{propfluc}
model.
We acknowledge the support of the PHAROS COST action (CA16214).
This research has made use of data obtained from the Chandra Data
Archive, and software provided by the Chandra X-ray Center (CXC) in
the application package \textsc{ciao}.
Partly based on observations obtained with {\it XMM-Newton}, an ESA
science mission with instruments and contributions directly funded by
ESA Member States and NASA.
This research has made use of data and software provided by the High
Energy Astrophysics Science Archive Research Center (HEASARC), which
is a service of the Astrophysics Science Division at NASA/GSFC.
MK acknowledges support from NWO through a Spinoza grant.
BDM acknowledges support from a Ram\'on y Cajal Fellowship
(RYC2018-025950-I).
ML acknowledges funding from the ERC under the EU’s Horizon 2020
research and innovation programme (consolidator grant agreement
No. 101002352).

\section{Data availability:}

The data underlying this article are available in HEASARC, at
https://heasarc.gsfc.nasa.gov/cgi-bin/W3Browse/w3browse.pl.

\bibliographystyle{mn2e}
\bibliography{../biblio.bib}

\end{document}